\documentclass[11pt]{article}

\usepackage{fullpage,amsmath,epsfig}
\usepackage{amssymb}

\begin{document}

\newtheorem{definition}{Definition}[section]
\newtheorem{theorem}{Theorem}[section]
\newtheorem{prop}[theorem]{Proposition}
\newtheorem{lemma}[theorem]{Lemma}
\newtheorem{remark}[theorem]{Remark}
\newtheorem{corollary}[theorem]{Corollary}
\newtheorem{observation}[theorem]{Observation}
\newtheorem{claim}[theorem]{Claim}
\newtheorem{conj}[theorem]{Conjecture}

\def\square{\vrule height6pt width7pt depth1pt}
\def\endpf{\hfill\square\bigskip}

\title{
{On Searching a Table Consistent with Division Poset
\bigskip} }

\date{} %Prevent the date from being printed

\author{\large{Yongxi Cheng${}^\mathrm{a,}$\thanks{Corresponding author.}\,\ , Xi Chen${}^\mathrm{a}$\,,
Yiqun Lisa Yin${}^\mathrm{b}$}\\
\\
${}^\mathrm{a}$\ \normalsize{Department of Computer Science, Tsinghua University, Beijing 100084, China}\\
\small{\tt \{cyx,xichen00\}@mails.tsinghua.edu.cn}\\
[1mm]
${}^\mathrm{b}$\ \normalsize{Independent security consultant, Greenwich CT, USA}\\
\small{\tt yiqun@alum.mit.edu} }

\maketitle

\begin{abstract}
Suppose $P_n=\{1,2,\ldots,n\}$ is a partially ordered set with the partial order defined by divisibility, that is,
for any two distinct elements $i,j\in P_n$ satisfying $i$ divides $j$, $i<_{P_n} j$. A table
$A_n=\{a_i|i=1,2,\ldots,n\}$ of distinct real numbers is said to be \emph{consistent} with $P_n$, provided for any
two distinct elements $i,j\in \{1,2,\ldots,n\}$ satisfying $i$ divides $j$, $a_i< a_j$. Given an real number $x$,
we want to determine whether $x\in A_n$, by comparing $x$ with as few entries of $A_n$ as possible. In this paper
we investigate the complexity $\tau(n)$, measured in the number of comparisons, of the above search problem. We
present a $\frac{55n}{72}+O(\ln^2 n)$ search algorithm for $A_n$ and prove a lower bound
$(\frac{3}{4}+\frac{17}{2160})n+O(1)$ on $\tau(n)$ by using an adversary argument.
\end{abstract}

\section{Introduction}
\label{sec:intro}

Suppose $P=\{1,2,\ldots,n\}$ is a partially ordered set (\emph{poset}), we say a table $A=\{a_i|i=1,2,\ldots,n\}$
of $n$ distinct real numbers \emph{consistent} with $P$, provided that for any $i, j\in P$ satisfying $i<_{P} j$,
$a_i< a_j$. Given a table $A$ of distinct real numbers which is consistent with a known poset $P$, and given a
real number $x$, we want to determine whether $x\in A$, by making a series of comparisons between $x$ and certain
elements $a_i\in A$. The problem is considered in a model using pairwise comparisons of the form $x : a_i$
($a_i\in A$) as basic operations. These comparisons have ternary outcomes $x<a_i$, $x=a_i$, or $x>a_i$. Our aim is
to make as few comparisons as possible. The \emph{complexity} of the problem is defined to be the minimum, over
all search algorithms for $P$, of the maximum number of comparisons required in the worst case.

In this article we consider the above search problem for the case where the partial order is defined by
divisibility. Let $P_n=\{1,2,\ldots,n\}$ be a poset with the partial order such that for any two distinct elements
$i,j\in P_n$ satisfying $i$ divides $j$, $i<_{P_n} j$. Thus we say a table $A_n=\{a_i|i=1,2,\ldots,n\}$ of
distinct real numbers consistent with $P_n$ if for any two distinct elements $i,j\in \{1,2,\ldots,n\}$ satisfying
$i$ divides $j$, $a_i< a_j$. Denote by $\tau(n)$ the complexity of the problem searching a given real number $x$
in $A_n$. We will investigate both upper bounds (i.e., search algorithms for $A_n$) and lower bounds on $\tau(n)$,
our main result is the following

\begin{theorem}\label{thm:main}
For $n\ge 1$, $c_1n+O(1)\le \tau(n) \le c_2n+O(\ln^2 n)$, where $c_1$ and $c_2$ are constants, and
$c_1=\frac{3}{4}+\frac{17}{2160}\approx 0.758$, $c_2=\frac{55}{72}\approx 0.764$.
\end{theorem}

Throughout the paper, we denote by $\lceil x \rceil$ (ceiling of $x$) the least integer that is not less than $x$,
denote by $\lfloor x \rfloor$ (floor of $x$) the largest integer that is not greater than $x$.

\section{Related Work}
\label{sec:relatedwork}

In~\cite{LS1, LS2} Linial and Saks studied the above class of search problems for general finite partially ordered
set $P$. In \cite{LS1} some general bounds on the complexity are provided, more precise results are also presented
for the case that $P$ is a product of chains and that $P$ is a rooted forest. In~\cite{LS2} they proved that, for
general finite partially ordered sets, the information theoretic bound for the complexity is tight up to a
multiplicative constant.

There are different perspectives on the problem of searching posets. In \cite{BGLY}, the authors studied the
tradeoff between the preprocessing time and the subsequent search time in a partial order. Let $P(n)$ be the
worst-case cost of a preprocessing algorithm which builds some partial orders, and let $S(n)$ be the maximum
number of comparisons required to answer a membership query. They proved that $P(n)+n\log_2 S(n)\ge (1+o(1))n
\log_2 n$ for any comparison-based algorithm.

A different notion of searching a poset was studied in~\cite{BFN, CDKL}. They consider searching a given element
$x$ in a poset $P$, instead of searching a given real number in a table of real numbers consistent with $P$. In
this case, for each comparison $x : p_i$ ($p_i\in P$) there are two possible outcomes: `yes' indicates that $x$ is
`below' $p_i$ (less than or equal to $p_i$); `no' indicates that $x$ is not below $p_i$. The aim is to find the
optimal search strategy. In spite of the similarity in definition, this turns out to be a quite different model.
In \cite{BFN}, the authors gave a polynomial time algorithm for posets having tree structures. In \cite{CDKL}, the
authors proved that the problem is NP-hard in general, they also gave a (1+o(1))-approximation algorithm under the
\emph{random graph} model and a 6.34-approximation algorithm under the \emph{uniform} model, both of these run in
polynomial time.

\section{Easy Bounds on $\tau (n)$}
\label{sec:easybounds}

First we give an easy lower bound $\frac{3n}{4}$ on $\tau (n)$ and a simple asymptotical $cn$ algorithm searching
$A_n$, where $c\approx 0.81$ is a constant.

\subsection{Lower Bounds $(3/4)n$}
\label{subsec:easylowerbound}

It is easy to see that the following simple \emph{response strategy} for the adversary can guarantee that at least
$\frac{3n}{4}$ comparisons of the form $x : a_i$ ($a_i \in A_n$) are required to determine whether $x\in A_n$, for
any search algorithm.
\\[1.5mm]
\indent \emph{\textbf{Response Strategy $RS_1$:}} \emph{when the algorithm asks about $x : a_i$,
$i=1,2,\ldots,\lfloor \frac{n}{2}\rfloor$, answer $x>a_i$; when the algorithm asks about $x : a_i$, $i=\lfloor
\frac{n}{2}\rfloor+1,\ldots,n$, answer $x<a_i$.}
\\[1.5mm]
\indent Focus on the subset $A^{*}_n=\{a_i|\frac{n}{4}<i\le n\}$ of $A_n$, if the adversary answers queries in
above way, the algorithm needs the comparison of $x$ with each element $a_i\in A^{*}_n$ to determine whether
$x=a_i$, thus needs at least $\frac{3n}{4}$ comparisons. In fact, the set $P^{*}_n=\{i \in P_n|\frac{n}{4}<i\le
n\}$ is a \emph{section} (see \cite{LS1}) of $P_n$ , and there is no ordered chain having length more than two in
$P^{*}_n$.

\subsection{A Simple Search Algorithm}

We will give an asymptotical $cn$ algorithm searching $A_n$ based on binary search, where
$c=\sum_{t=0}^{\infty}\frac{1}{2^{2^t}}\approx 0.81$ is a constant. Define $B_i=\{a_j\in A_n|j=(2i-1)\times 2^k \
(\le n),\ k=0, 1,\ldots\}$, $i=1,2,\ldots,\lceil n/2\rceil$. Then $B_1$, $B_2$,\ldots,$B_{\lceil n/2\rceil}$ is a
partition of $A_n$, and each $B_i$ is in linear ordering. The algorithm runs as follows.
\\[1.5mm]
\noindent \textit{\textbf{Algorithm 1 }}\emph{(Searching Table $A_n$):} \ Binary search $B_1$,
$B_2$,\ldots,$B_{\lceil n/2\rceil}$ one by one.
\\[1.5mm]
\indent Now we analyze the number of comparisons required by Algorithm~1. For each $B_i$, binary search needs at
most $\lceil \log_2 (|B_i|+1) \rceil$ comparisons, $i=1,2,\ldots,\lceil n/2\rceil$. Since $1\leq |B_i| \leq
\lfloor \log_2 n\rfloor +1$, and the number of the sets $B_i$ having exactly $k$ elements is less than
$n/2^{k+1}+1$, it follows that the total number of comparisons required by Algorithm~1 is at most
\begin{eqnarray*}
\begin{aligned}
s_1(n)&=\sum_{k=1}^{\lfloor \log_2 n\rfloor +1} (\frac{n}{2^{k+1}}+1) \lceil \log_2 (k+1) \rceil\\
    &<\sum_{k=1}^{\infty} \frac{n}{2^{k+1}} \lceil \log_2 (k+1) \rceil+\sum_{k=1}^{\lfloor \log_2 n\rfloor +1} \lceil \log_2 (k+1) \rceil\\
    &=n\cdot \sum_{t=0}^{\infty}\frac{1}{2^{2^t}}+O(\ln n \cdot \ln\ln n)
\end{aligned}
\end{eqnarray*}

\section{Upper Bounds $\frac{55n}{72}+O(\ln^2 n)$ on $\tau (n)$}
\label{sec:upperbound}

In this section we present a $\frac{55n}{72}+O(\ln^2 n)$ search algorithm for $A_n$, by partitioning $A_n$ into
2-dimensional \emph{layers} and then searching them one by one.

Let $I_n=\{i|1\le i\le n, \ \mbox{$i$ does not have factor 2 or 3}\}$. For each $i\in I_n$, extend $a_i$ to a
subset of $A_n$, $L_i$, such that $L_i=\{a_{i\times 2^k\times 3^s}\in A_n|k,s=0,1,2,\ldots\}$. For instance, for
$n=15$, we have subsets $L_1$, $L_5$, $L_7$, $L_{11}$, $L_{13}$, and
\begin{eqnarray*}
L_1=\left[
\begin{array}{ccccc}
 a_1& a_2& a_4  & a_8    \\
 a_3& a_6& a_{12} &      \\
 a_9&  &    &      \\
\end{array}%
\right]\ \ L_5=\left[
\begin{array}{ccccc}
a_5&a_{10} \\
a_{15}&  \\
\end{array}%
\right]\ \ L_7=\left[
\begin{array}{ccccc}
a_7&a_{14}\\
\end{array}%
\right]\ \ L_{11}=\left[
\begin{array}{ccccc}
a_{11}\\
\end{array}%
\right]\ \ L_{13}=\left[
\begin{array}{ccccc}
a_{13}\\
\end{array}%
\right]
\end{eqnarray*}
It is easy to see that all these subsets $L_i$ (which sometimes will be referred to as \emph{layers}), $i\in I_n$,
form a partition of $A_n$.

An algorithm searching an real number $x$ in an $m\times n$ ($m,n\geq 1$) \emph{monotone matrix} (a matrix with
entries increasing along each row and each column) was described in~\cite{LS1}, which repeats comparing $x$ with
the element $e$ at the top right corner of the current matrix, either the first row or the rightmost column of the
current matrix will be eliminated depending on whether $x>e$ or $x<e$, thus the algorithm requires at most $m+n-1$
comparisons (see \cite{GK, LS1} for the lower bounds on the number of comparisons required for this problem when
$m=n$). Based on this algorithm and notice that a layer is a ``triangular'' portion of a monotone matrix, we can
apply the above ``$m+n-1$'' algorithm searching a layer. Furthermore, for some of the layers we can do slightly
better by exploiting the properties of the layers.

\begin{lemma}
\label{lemma:searchlayer} If a layer $L$ with $|L|\notin \{1,2,3,5\}$, then one can search $x$ in $L$ using at
most $m+n-2$ comparisons, where $m$ and $n$ are the numbers of rows and columns of $L$, respectively.
\end{lemma}

\noindent \textbf{Proof.} See Appendix \ref{appendix:searchlayer}. \endpf

\noindent By Lemma \ref{lemma:searchlayer} we can obtain an improved search algorithm for $A_n$.
\\
\\
\noindent \textit{\textbf{Algorithm 2}} \emph{(Searching Table $A_n$):}
\\[1mm]
\indent Search all layers $L_i$ one by one. If $|L_i|\in \{1,2,3,5\}$, search $L_i$ using the $``m+n-1"$ algorithm
in \cite{LS1}; Otherwise, search $L_i$ using the $``m+n-2"$ algorithm in Lemma \ref{lemma:searchlayer}.
\\
\\
Next we analyze the number of comparisons required by Algorithm~2. Define $r(L_i)$ and $c(L_i)$ to be the numbers
of rows and columns of $L_i$ respectively. The total number of comparisons required by Algorithm~2 is at most
\begin{eqnarray*}
\begin{aligned}
s_2(n)&=\sum_{i\in I_n,\ |L_i|\in \{1,2,3,5\}}^{} (r(L_i)+c(L_i)-1) \ +\sum_{i\in I_n,\ |L_i|\notin \{1,2,3,5\}}^{} (r(L_i)+c(L_i)-2)\\
    &=\sum_{i\in I_n}^{} r(L_i)+\sum_{i\in I_n}^{} c(L_i)-2\sum_{i\in I_n}^{} 1+\sum_{i\in I_n,\ |L_i|\in \{1,2,3,5\}}^{} 1
\end{aligned}
\end{eqnarray*}

Clearly, $r(L_i)\le 1+\log_3 n$ for any $i\in I_n$. The number of layers $L_i$ with $r(L_i)=p$ is less than
$\frac{2n}{3^{p+1}}+1$, since $3^{p-1}i\leq n$, $3^p i>n$ and $i$ has no factor 2 or 3. Similarly, $c(L_i)\le
1+\log_2 n$ for any $i\in I_n$, and the number of layers $L_i$ with $c(L_i)=q$ is less than $\frac{n}{3\times
2^q}+1$. It follows that
\begin{eqnarray*}
\begin{aligned}
&\sum_{i\in I_n} r(L_i)<\sum_{p=1}^{1+\log_3 n}(\frac{2n}{3^{p+1}}+1)
p<\sum_{p=1}^{\infty}\frac{2np}{3^{p+1}}+\sum_{p=1}^{1+\log_3 n}p=\frac{n}{2}+O(\ln^2 n),\\
&\sum_{i\in I_n} c(L_i)<\sum_{q=1}^{1+\log_2 n}(\frac{n}{3\times 2^q}+1) q<\sum_{q=1}^{\infty}\frac{nq}{3\times
2^q}+\sum_{q=1}^{1+\log_2 n} q=\frac{2n}{3}+O(\ln^2 n).
\end{aligned}
\end{eqnarray*}
In addition, $\sum_{i\in I_n}^{} 1=|I_n|=\frac{n}{3}+O(1)$, and

\begin{eqnarray*}
\begin{aligned}
 \sum_{i\in I_n,\ |L_i|\in \{1,2,3,5\}}^{} 1=\sum_{i\in I_n, \,\frac{n}{2}<i\le n }^{} 1 + \sum_{i\in I_n, \,\frac{n}{3}<i\le \frac{n}{2} }^{}
 1 + \sum_{i\in I_n, \,\frac{n}{4}<i\le \frac{n}{3} }^{} 1 + \sum_{i\in I_n, \,\frac{n}{8}<i\le \frac{n}{6} } ^{}
 1 =\frac{19n}{72}+O(1).
\end{aligned}
\end{eqnarray*}
Therefore, $s_2(n)\le \frac{55n}{72}+O(\ln^2 n)$.

%=&\frac{1}{3}(n-\frac{n}{2})+O(1)+\frac{1}{3}(\frac{n}{2}-\frac{n}{3})+O(1)+\frac{1}{3}(\frac{n}{3}-\frac{n}{4})+O(1)+\frac{1}{3}(\frac{n}{6}-\frac{n}{8})+O(1)\\

\section{Lower Bounds $(\frac{3}{4}+\frac{1}{432})n+O(1)$ on $\tau (n)$}
\label{sec:lowerbound1}

In this section we prove a lower bound $(\frac{3}{4}+\frac{1}{432})n+O(1)$ on $\tau (n)$,  by using an adversary
argument. Thus the previous easy lower bound $\frac{3n}{4}$ is not best possible.

\subsection{The Main Idea in Constructing Lower Bounds}
\label{subsec:mainidea}

Recall the response strategy $RS_1$ for the adversary given in Section~\ref{sec:easybounds}, which guarantees that
at least $\frac{3n}{4}$ comparisons are needed to determine whether $x\in A_n$ for any search algorithm. We can
view $RS_1$ in the following way.

In Algorithm~2, $A_n$ is partitioned into layers $L_i$. For each row $R$ in each layer, if $R$ has only one
element, we pick this element and say that it forms a \emph{unit}; if $R$ has at least two elements, we pick the
last two elements as a unit. We will call a unit consisting of one or two elements \emph{1-unit} or \emph{2-unit},
respectively. In total $\frac{3n}{4}$ elements are picked, which form exactly the subset
$A^{*}_n=\{a_i|\frac{n}{4}<i\le n\}$ of $A_n$. We can now restate response strategy $RS_1$ as follows.
\\
\\
\noindent \emph{response strategy for the elements in a 1-unit:} For $\frac{n}{2}<i\le n$, $i$ is odd, when the
algorithm asks about $x : a_i$, answer $x<a_i$.
\begin{equation*}
\left[
\begin{array}{ccccccccc}
 \ldots               &\ldots\\
                      &      \\
\{\ \ulcorner a_i \ \}&      \\
\end{array}
\right]
\end{equation*}
\\
%---------------------------------------------2-unit:-------------------------------------------
\emph{response strategy for the elements in a 2-unit:} For $\frac{n}{4}<i\le \frac{n}{2}$, when the algorithm asks
about $x : a_i$, answer $x>a_i$; when the algorithm asks about  $x : a_{2i}$, answer $x<a_{2i}$.
\begin{equation*}
\left[
\begin{array}{ccccccccc}
\ldots& \ldots             &\ldots              &\ldots\\
      &                    &                    &      \\
\ldots&\{\ a_i \lrcorner\ ,&\ulcorner a_{2i}\ \}&      \\
      &                    &                    &      \\
\ldots& \ldots             &                    &      \\
\end{array}
\right]
\end{equation*}
\\
\emph{response strategy for the elements do not belong to any unit:} For $1 \le i\le \frac{n}{4}$, when the
algorithm asks about $x : a_i$, answer $x>a_i$.

In general, if an algorithm compares $x$ with $a_i$ and gets the result $x<a_i$, then all the elements $a_k$ with
$k$ divisible by $i$ are known to be larger than $x$, thus could be eliminated from consideration, we say that
these elements are \emph{cut} by $a_i$. Similarly, if the algorithm gets the result $x>a_i$, then all the elements
$a_k$ with $k$ that divides $i$ are known to be smaller than $x$ and could be eliminated, we also say that these
elements are cut by $a_i$.

It is easy to see that under response strategy $RS_1$, any element that belongs to some unit could not be cut by
any other element. Therefore, if strategy $RS_1$ is adopted by the adversary, for any search algorithm in order to
determine whether $x\in A_n$, the $\frac{3n}{4}$ comparisons of $x$ with all the elements in all the units are
necessary, thus obtaining the lower bound $\frac{3n}{4}$.

Hereafter, in some response strategy, the notation `$a_i \lrcorner$' means that when queried by any algorithm the
comparison $x : a_i$, answer $x>a_i$, and we say that \emph{$a_i$ cuts to the left up}; `$\ulcorner a_i$' means
that when queried the comparison $x : a_i$, answer $x<a_i$, and we say that \emph{$a_i$ cuts to the right bottom}.
\\
\\
\indent A natural thought for constructing better lower bounds could be: if more elements from each row are
picked, can we prove that more than $\frac{3n}{4}$ comparisons are required? Picking three elements from a row
makes no difference, since all elements in a row are in linear order, and two comparisons are sufficient for
searching three ordered elements. Therefore, at least four elements should be chosen from some rows, to guarantee
that at least three comparisons are required to search $x$ in them.

We will pick elements in the following way. If there are less than four elements in a row, we pick all the
elements as a unit; if there are at least four elements in a row, we pick the last four elements as a unit.
However, in general, we could no longer guarantee, as we do under strategy $RS_1$, that each of the elements
picked can not be cut by other elements. Actually we even could not guarantee that an element picked can not be
cut by an element in other units. An element that can not be cut by any element outside its unit, under some
response strategy, can guarantee the number of comparisons required by any search algorithm. We call this kind of
elements \emph{essential} (notice that an essential element may be cut by other elements in the same unit). Next
we will present a more effective response strategy, in which there are sufficient essential elements to guarantee
that more than $\frac{3n}{4}$ comparisons are required, for any search algorithm.

\subsection{Units and Special Units}
\label{subsec:units}

Let us start with some definitions. As described above, there is exactly one \emph{unit} in each row of each
layer. If in a row there are less than four elements, all these elements form a unit; if in a row there are at
least four elements, the last four elements form a unit. A unit consisting of one, two, three, or four elements is
called \emph{1-unit}, \emph{2-unit}, \emph{3-unit}, or \emph{4-unit} respectively. E.g., in the below layer
(\ref{R2:4-unit-s}), we have a 1-unit $\{a_{9i}\}$, a 3-unit $\{a_{3i}, a_{6i}, a_{12i}\}$, and a 4-unit
$\{a_{2i}, a_{4i}, a_{8i}, a_{16i}\}$.

Next we introduce an important subcollection of the units defined above, called \emph{special units}, which is the
key to the proof of new lower bounds.
\begin{definition}
\label{def:specialunit} \emph{\emph{special units:} A unit $u$ is called a \emph{special unit} if $u$ is the
4-unit in a layer $L_i$ with $|L_i|=9$ (i.e., a layer $L_i$ with $i\in S_n$, where $S_n=\{i|\frac{n}{18}<i\le
\frac{n}{16}$, $i$ is not divisible by 2 or 3$\}$). We denote by 4-$unit_s$ a special unit.}
\end{definition}

Since the form of a layer $L_i$ is determined by its number of elements, $|L_i|$, a layer containing a special
unit must have the following form (the marks `$\lrcorner$' and `$\ulcorner$' indicating the cut directions of the
elements will be explained later in the new response strategy, $RS_2$). Each such layer contains exactly one
special unit, $\{a_{2i}, a_{4i}, a_{8i}, a_{16i}\}$, in its first row.
\begin{equation} \label{R2:4-unit-s}
\left[
\begin{array}{ccccccccc}
a_i\ ,          &\{\ a_{2i}\lrcorner\ ,&\ulcorner a_{4i}\lrcorner\ ,&\ulcorner a_{8i}\lrcorner\ ,&\ulcorner a_{16i}\ \}\\
                &                  &                        &                        &                 \\
a_{3i}\lrcorner\ ,&\ulcorner a_{6i}\ ,   &\ulcorner a_{12i}           &                        &                 \\
              &                  &                        &                        &                 \\
 \ulcorner a_{9i} &                  &                        &                        &                 \\
\end{array}
\right]
\end{equation}

A unit that is not a special unit is called a \emph{general unit}, thus all 1-units, 2-units, and 3-units are
general units, a subcollection of 4-units are general units.

Special units can help proving better lower bounds because we will prove later that, under response strategy
$RS_2$ described below, all elements in special units are \emph{essential}, thus each special unit guarantees at
least three necessary comparisons for any search algorithm, which is one more than the two necessary comparisons
guaranteed by the 2-unit in its row under $RS_1$, consequently the lower bound $\frac{3n}{4}$ can be improved by
the number of special units, $|S_n|$.

\subsection{New Response Strategy}

Now we are ready to describe the new response strategy for the adversary, and show that which can guarantee that
at least $(\frac{3}{4}+\frac{1}{432})n+O(1)$ comparisons are required for any search algorithm to determine
whether $x\in A_n$.
\\
\\
\emph{\textbf{Response Strategy $RS_2$:}}

\indent \emph{for the elements not in a unit:} For $1 \le i\le \frac{n}{16}$, when the algorithm asks about $x :
a_i$, answer $x>a_i$.

%---------------------------------------------1-unit:-------------------------------------------
\indent \emph{for the elements in a 1-unit:} For $\frac{n}{2}<i\le n$, $i$ is odd, when the algorithm asks about
$x : a_i$, answer $x<a_i$.
\begin{equation} \label{R2:1-unit}
\left[
\begin{array}{ccccccccc}
 \ldots             &\ldots\\
                    &      \\
\{\ \ulcorner a_i \ \}&      \\
\end{array}
\right]
\end{equation}

%---------------------------------------------2-unit:-------------------------------------------
\emph{for the elements in a 2-unit:} For $\frac{n}{4}<i\le \frac{n}{2}$, $i$ is odd, when the algorithm asks about
$x : a_i$, answer $x>a_i$; when the algorithm asks about $x : a_{2i}$, answer $x<a_{2i}$.
\begin{equation} \label{R2:2-unit}
\left[
\begin{array}{ccccccccc}
 \ldots           &\ldots          &\ldots\\
                  &                &      \\
\{\ a_i \lrcorner\ ,&\ulcorner a_{2i}\ \}&      \\
                  &                &      \\
 \ldots           &                &      \\
\end{array}
\right]
\end{equation}
\indent The 3-units are partitioned into two classes according to the number of elements of the next row below
them, which are denoted by 3-$unit_1$'s and 3-$unit_2$'s respectively.
%---------------------------------------------3-$unit_1$-------------------------------------------

\emph{for the elements in a 3-$unit_1$ (the 1st class of 3-units whose next row contains one element):} For
$\frac{n}{6}<i \le \frac{n}{4}$, $i$ is odd, when the algorithm asks about $x : a_i$, answer $x>a_i$; when the
algorithm asks about $x : a_{2i}$ or $x : a_{4i}$, answer $x<a_{2i}$ or $x<a_{4i}$ respectively.
\begin{equation} \label{R2:3-unit-1}
\left[
\begin{array}{ccccccccc}
 \ldots            &\ldots         &\ldots             &\ldots\\
                   &               &                   &      \\
\{\ a_{i} \lrcorner \ ,&\ulcorner a_{2i}\ ,&\ulcorner  a_{4i} \ \} &      \\
                   &               &                   &      \\
\  \ulcorner a_{3i}\   &               &                   &      \\
\end{array}
\right]
\end{equation}

%---------------------------------------------3-$unit_2$-------------------------------------------
\emph{for the elements in a 3-$unit_2$ (the 2nd class of 3-units whose next row contains two elements):} For
$\frac{n}{8}<i\le \frac{n}{6}$, $i$ is odd, when the algorithm asks about $x : a_i$ or $x : a_{2i}$, answer
$x>a_i$ or $x>a_{2i}$ respectively; when the algorithm asks about $x : a_{4i}$, answer $x<a_{4i}$.
\begin{equation} \label{R2:3-unit-2}
\left[
\begin{array}{ccccccccc}
 \ldots            &\ldots          &\ldots              &\ldots\\
                   &                &                    &      \\
\{\  a_{i} \lrcorner\ ,&a_{2i} \lrcorner \ ,&\ulcorner a_{4i}\ \}    &      \\
                   &                &                    &      \\
  a_{3i}\lrcorner   \ ,&  \ulcorner a_{6i}  &                    &      \\
\end{array}
\right]
\end{equation}
\indent There are two classes of 4-units, general 4-units and special 4-units, which are denoted by 4-$unit_g$'s
and 4-$unit_s$'s respectively.

%---------------------------------------------4-$unit_g$-------------------------------------------
\emph{for the elements in a 4-$unit_g$ (general 4-unit):} For $\frac{n}{16}<i\le \frac{n}{8}$, $i\notin 2S_n$
(where $2S_n=\{2j|j\in S_n\}$, and see Definition \ref{def:specialunit} for $S_n$), when the algorithm asks about
$x : a_i$ or $x : a_{2i}$, answer $x>a_i$ or $x>a_{2i}$ respectively; when the algorithm asks about $x : a_{4i}$
or $x : a_{8i}$, answer $x<a_{4i}$ or $x<a_{8i}$ respectively.
\begin{equation} \label{R2:4-unit-c}
\left[
\begin{array}{ccccccccc}
\ldots&\ldots              &\ldots         &  \ldots      &\ldots            &\ldots\\
      &                    &               &              &                  &      \\
\ldots&\{\  a_i \lrcorner\ , &a_{2i}\lrcorner\ , &\ulcorner a_{4i}\ ,&\ulcorner  a_{8i} \ \}&      \\
      &                    &               &              &                  &      \\
\ldots&\ldots              &\ldots         &  \ldots      &                  &      \\
\end{array}
\right]
\end{equation}

%---------------------------------------------4-$unit_s$-------------------------------------------*******************
\emph{for the elements in a 4-$unit_s$ (special 4-unit):} For $i\in S_n$, $\{a_{2i}, a_{4i}, a_{8i}, a_{16i}\}$ is
a special unit. The response strategy for elements in a special unit is adaptive, depending on the order of
comparisons with $x$ made by the search algorithm, to guarantee that at least three comparisons are needed to
determine whether $x\in \{a_{2i}, a_{4i}, a_{8i}, a_{16i}\}$.

\noindent -- If the algorithm first asks about $x : a_{2i}$, answer $x>a_{2i}$. Then $a_{2i}$ will be eliminated,
and the remained three elements will follow the strategy $\{a_{4i} \lrcorner, \ulcorner a_{8i}, \ulcorner
a_{16i}\}$ for possible subsequent comparisons with $x$.

\noindent -- If the algorithm first asks about $x : a_{4i}$, answer $x>a_{4i}$. Then $a_{2i}$ and $a_{4i}$ are
known to be smaller than $x$ and will be eliminated, the remained two elements will follow the strategy
$\{a_{8i}\lrcorner, \ulcorner a_{16i}\}$.

\noindent -- If the algorithm first asks about $x : a_{8i}$, answer $x<a_{8i}$. Then $a_{8i}$ and $a_{16i}$ will
be eliminated, and the remained two elements will follow the strategy $\{a_{2i}\lrcorner, \ulcorner a_{4i}\}$.

\noindent -- If the algorithm first asks about $x : a_{16i}$, answer $x<a_{16i}$. Then $a_{16i}$ will be
eliminated, and the remained three elements will follow the strategy $\{a_{2i}\lrcorner, a_{4i} \lrcorner,
\ulcorner a_{8i}\}$.
\\
We denote the response strategy of elements in a special unit by $\{a_{2i}\lrcorner, \ulcorner a_{4i}\lrcorner,
\ulcorner a_{8i}\lrcorner, \ulcorner a_{16i}\}$, see (\ref{R2:4-unit-s}).

\subsection{Lower Bounds $(\frac{3}{4}+\frac{1}{432})n+O(1)$ on $\tau(n)$}
\label{subsec:lowerbound1}
%-----------------------------------------------------further revision------------------------------------------

Recall that we call an element $a_i \in A_n$ \emph{essential} under strategy $RS_2$, if $a_i$ belongs to some unit
$u$ (i.e., $i>\frac{n}{16}$), and any element not in $u$ can not cut $a_i$ under $RS_2$.
\\
\\
Define set \emph{$E_n$=\{all elements of 1-units\} $\cup$ \{all elements of 2-units\} $\cup$ \{all the first and
second elements of 3-$unit_1$'s\} $\cup$ \{all the second and third elements of 3-$unit_2$'s\} $\cup$ \{all the
second and third elements of 4-$unit_g$'s\} $\cup$ \{all elements of 4-$unit_s$'s\}.} We can prove the following
lemma.

\begin{lemma}
\label{essential1} Under response strategy $RS_2$, all elements of $E_n$ are essential.
\end{lemma}

\noindent \textbf{Proof of Lemma \ref{essential1}.} See Appendix \ref{appendix:essential1}. \endpf

By Lemma \ref{essential1} and response strategy $RS_2$, each essential element in general units needs one
comparison with $x$ to determine whether it equals $x$. For each special unit, at least three comparisons between
$x$ and its elements are needed to determine whether $x$ is in it. By comparing the 1-units and 2-units picked for
$RS_1$ in Section~\ref{subsec:mainidea}, we can see that under strategy $RS_2$, each row containing a general unit
contributes the same number of necessary comparisons as it contributes under $RS_1$. While for each row containing
a special unit, it contributes one more necessary comparison than it does under $RS_1$. Therefore the lower bound
$\frac{3n}{4}$ could be improved by the number of special units, $|S_n|$, which is $|\{i: \frac{n}{18}<i\le
\frac{n}{16}$, $i$ is not divisible by 2 or 3$\}|$=$(\frac{n}{16}-\frac{n}{18})\times
\frac{1}{3}+O(1)=\frac{n}{432}+O(1)$. Thus we get a new lower bound $(\frac{3}{4}+\frac{1}{432})n+O(1)$.

\section{Improved Lower Bounds $(\frac{3}{4}+\frac{17}{2160})n+O(1)$ on $\tau(n)$}
\label{sec:lowerbound2}

By recognizing more special units, we can extend the above method to obtain a better lower bound
$(\frac{3}{4}+\frac{17}{2160})n+O(1)$ on $\tau(n)$. The units are defined in the same way as in
Section~\ref{subsec:units}, and we will introduce three classes of special units.

\begin{definition}
\emph{\emph{the 1st class of special units (4-$unit_{s,1}$):} A unit $u$ is of the 1st class of special units if
$u$ is the 4-unit in a layer $L_i$ with $|L_i|=9$ and $i$ is not divisible by 5.}
\end{definition}

The definition of 4-$unit_{s,1}$ is just Definition \ref{def:specialunit} of 4-$unit_{s}$ with an extra
restriction that $i$ is not divisible by 5. A layer $L_i$ containing a 4-$unit_{s,1}$ must have the form
(\ref{R2:4-unit-s}), each such layer contains exactly one special unit $\{a_{2i}, a_{4i}, a_{8i}, a_{16i}\}$ in
its first row. The subscripts $j=2i$ of the first elements of all 4-$unit_{s,1}$'s form a set $S_{n,1}=\{j
(=2i)|\frac{n}{9}<j\le \frac{n}{8}$, $j$ is divisible by 2, but is not divisible by 3, 4 or 5$\}.$

\begin{definition}
\emph{\emph{the 2nd class of special units (4-$unit_{s,2}$):} A 4-unit $u$ is of the 2nd class of special units if
it has the following properties:}
\begin{enumerate}
\item \emph{$u$ is the 4-unit in the \emph{first} row of a layer $L_j$ with $j$ not divisible by 5.}

\item \emph{The first row of $L_j$ has at least six elements and has two more elements than the second row of $L_j$.}
\end{enumerate}
\end{definition}

The layers containing a 4-$unit_{s,2}$ must have the following form (\ref{specialunit-2}), where the symbol
`$\bullet$' indicates that there must exist an element at the position. Each such layer contains exactly one
special unit, $\{a_{i}, a_{2i}, a_{4i}, a_{8i}\}$, in its first row. The subscripts $i$ of the first elements of
all 4-$unit_{s,2}$'s form a set $S_{n,2}=\{i|\frac{n}{12}<i\le \frac{n}{8}$, $i$ is divisible by 4, but is not
divisible by 3 or 5$\}.$
\begin{equation}
\label{specialunit-2} \left[
\begin{array}{ccccccccccc}
\ldots  &\bullet     &\bullet       &\{\ a_{i} \lrcorner  &\ulcorner a_{2i} \lrcorner    &\ulcorner a_{4i} \lrcorner     &\ulcorner a_{8i} \}\\
       &            &              &                    &          &        &                      \\
\ldots  &\ \bullet \lrcorner  &\bullet \lrcorner &   \ulcorner a_{3i}   &\ulcorner a_{6i} &                &             \\
\      &            &              &                    &          &        &                      \\
\ldots  &\ldots       &\ldots         &\ldots               &          &        &                      \\
\end{array}
\right]
\end{equation}

\begin{definition}
\emph{\emph{the 3rd class of special units (4-$unit_{s,3}$):} A 4-unit $u=\{a_{i}, a_{2i}, a_{4i}, a_{8i}\}$ is of
the 3nd class of special units if it is in a layer $L_j$ with $j$ not divisible by 5, and has the following
properties:}
\begin{enumerate}
\item \emph{The subscript of its first element, $i$, is divisible by 36 (i.e., in $L_j$, there are at least two rows above
$u$, and at least two columns before the first element $a_{i}$ of $u$).}

\item \emph{Denote by $R$ the row containing $u$ in $L_j$. The next row above $R$ has two more elements than $R$,
and the next row below $R$ has two less elements than $R$.}
\end{enumerate}
\end{definition}

A 4-$unit_{s,3}$ in a layer must have the following form (\ref{specialunit-3}) (see Lemma
\ref{lemma:increment_two} in Appendix \ref{appendix:essential2}), here we only list the subscripts.

\begin{equation}
\label{specialunit-3} \left[
\begin{array}{ccccccccccccc}
\ldots&\ldots &\ldots   &\ldots       & \ldots &\ldots   &\ldots   &\ldots &\ldots    & \ldots   &\ \ \ldots   \\
&&&           &        &          &        &    &        &    &   &\\
\ldots&\ldots   &\ldots   &\ldots       & \ldots &\ldots   &\bullet  &\bullet &\bullet & \bullet   &  \\
&&&           &        &          &        &    &        &    &   &\\
\ldots&\ldots &\ldots& \ldots  &\ldots &\frac{4i}{3} \lrcorner   &\frac{8i}{3} \lrcorner    &\ulcorner \frac{16i}{3} &\ulcorner \frac{32i}{3}&    &   \\
&&&           &        &          &        &    &        &    &   &\\
\ldots&\ldots &\ldots &\{\ i \lrcorner  &\ulcorner 2i \lrcorner    &\ulcorner 4i \lrcorner     &\ulcorner 8i\}&    &           &   &\\
&&&           &        &          &        &    &        &    &   &\\
\ldots& \bullet \lrcorner&\bullet \lrcorner&\ulcorner 3i     &\ulcorner 6i &          &        &    &        &       &\\
&&&           &        &          &        &    &        &    &   &\\
\ldots&\bullet   &\bullet  &\ulcorner 9i  &        &          &        &    &        &    &\\
&&&           &        &          &        &    &        &    &   &\\
\ldots&\ldots  &   &     &        &          &        &    &        &    &\\
\end{array}
\right]
\end{equation}
The subscripts $i$ of the first elements of all 4-$unit_{s,3}$'s form a set $S_{n,3}=\{i|\frac{n}{12}<i\le
\frac{3n}{32}$, $i$ is divisible by 4 and 9, but is not divisible by 5$\}.$
\\
\\
Based on the above new defined special units, we have the following response strategy.
\\
\\
\indent \emph{\textbf{Response Strategy $RS^{*}_2$:}}  In $RS^{*}_2$, the response strategy for the elements in
general units, in special units and not in any units are the same as their strategy in $RS_2$ respectively, with
only a different partition of 4-units into general units and special units.
\\
\\
\indent Similarly define set \emph{$E^{*}_n$=\{all elements of 1-units\} $\cup$ \{all elements of 2-units\} $\cup$
\{all the first and second elements of 3-$unit_1$'s\} $\cup$ \{all the second and third elements of 3-$unit_2$'s\}
$\cup$ \{all the second and third elements of 4-$unit_g$'s\} $\cup$ \{all elements of special units
4-$unit_{s,1}$'s, 4-$unit_{s,2}$'s and 4-$unit_{s,3}$'s\}.} We can prove the following lemma.

\begin{lemma}
\label{essential2} Under $RS^{*}_2$, all elements of $E^{*}_n$ are essential.
\end{lemma}

\noindent \textbf{Proof of Lemma \ref{essential2}.} See Appendix \ref{appendix:essential2}. \endpf

Using similar arguments as in Section~\ref{subsec:lowerbound1}, the lower bound $\frac{3n}{4}$ could be improved
by the number of the new defined special units, $|S_{n,1}|+|S_{n,2}|+|S_{n,3}|$, which is
\begin{eqnarray*}
\begin{aligned}
 &(\frac{n}{8}-\frac{n}{9})\times \frac{1}{4} \times \frac{2}{3}\times
 \frac{4}{5}+O(1)+(\frac{n}{8}-\frac{n}{12})\times \frac{1}{4}\times \frac{2}{3}\times
 \frac{4}{5}+O(1)+(\frac{3n}{32}-\frac{n}{12})\times \frac{1}{4}\times
 \frac{1}{9}\times \frac{4}{5}+O(1)\\
=&\frac{17n}{2160}+O(1).
\end{aligned}
\end{eqnarray*}
Therefore, we obtain an improved lower bound $(\frac{3}{4}+\frac{17}{2160})n+O(1)$.

\section{Concluding Remarks and Open Problems}
\label{sec:concluding}

In this paper we investigate the complexity, $\tau(n)$, of the problem searching a table consistent with division
poset. Our main result is the following. For $n\ge 1$, $c_1n+O(1)\le \tau(n) \le c_2n+O(\ln^2 n)$, where $c_1$ and
$c_2$ are constants, and $c_1=\frac{3}{4}+\frac{17}{2160}\approx 0.758$, $c_2=\frac{55}{72}\approx 0.764$. It may
be of interest to further close the gap.

Notice that under the model in this paper we only allow comparisons of the form $x : a_i$, i.e., all comparisons
must involve $x$. If we also allow pairwise comparisons among elements of $A_n$, then the techniques used in this
paper to prove lower bounds will not apply directly. It may be interesting to investigate the complexity of the
search problem under this new model.

\section*{Acknowledgments}

The authors are grateful to Andy Yao for introducing this interesting problem and insightful comments, to Xiaoming
Sun and Chen Wang for helpful discussions.

%\newpage
\

\noindent \textbf{\LARGE {Appendices}}

\begin{appendix}

\section{Proof of Lemma \ref{lemma:searchlayer}}

\label{appendix:searchlayer}

It is easy to see that the form of $L$ is determined by its cardinality $|L|$. First we prove two lemmas that will
be useful later.

\begin{lemma} \label{lemma:L1} In any layer, the difference of the lengths of any two
consecutive rows must be 1 or 2.
\end{lemma}

\noindent \textbf{Proof of Lemma~\ref{lemma:L1}.} We prove the lemma by contradiction. Otherwise, at least one of
the following two situations exists.
\begin{equation*}
\left[
\begin{array}{ccccc}
& \ldots & a_i   &       &     \\
& \ldots &a_{3i} &       &     \\
\end{array}%
\right]\ \ \ \ \ \ \ \ \ \ \ \ \ \ \ \ \ \ \ \ \left[
\begin{array}{cccccccccc}
& \ldots & a_i   &a_{2i}&a_{4i}&a_{8i}& \ldots &\\
& \ldots &a_{3i}&       &       &       &       &\\
\end{array}%
\right]
\end{equation*}
However, according to the definition of layers, the left one can not happen because the element $a_{2i}$ should be
in the layer, and the right one can not happen because the element $a_{6i}$ should be in the layer. Thus the lemma
holds. \endpf

\begin{lemma} \label{lemma:L2}
In any layer, there can not be three consecutive rows with lengths each increased by one.
\end{lemma}

\noindent \textbf{Proof of Lemma~\ref{lemma:L2}.} Otherwise, it must be the following situation.
\begin{equation*}
\left[
\begin{array}{cccccccccc}
& \ldots & a_i   &a_{2i}&a_{4i}&       & \\
& \ldots &a_{3i}&a_{6i}&       &       & \\
& \ldots &a_{9i}&       &       &       & \\
\end{array}%
\right]
\end{equation*}
However it can not happen since the element $a_{8i}$ should be in the layer. The lemma holds. \endpf

\noindent Now we are ready to prove Lemma \ref{lemma:searchlayer}. By Lemma \ref{lemma:L1}, we have the following
two cases.
\\[2.5mm]
\noindent \emph{Case 1.} The first row of $L$ has two more elements than the second row. Since $|L|\geq 4$, $L$
must have the following form

\begin{equation*}
\left[
\begin{array}{ccccccc}
& \ldots &  a_i    & a_{2i} & a_{4i} & \\
& \ldots & a_{3i} &         &         & \\
& \ldots &         &         &         & \\
\end{array}%
\right]
\end{equation*}
First compare $x$ with $a_{2i}$.

If $x<a_{2i}$, then $a_{2i}$ and $a_{4i}$ are known to be greater than $x$, and will be eliminated from $L$,
leaving a portion of an $m\times (n-2)$ monotone matrix, which can be searched using at most $m+(n-2)-1$
comparisons. In total, at most $m+n-2$ comparisons are needed.

If $x>a_{2i}$, then all elements in the first row except $a_{4i}$ are known to be smaller than $x$, and will be
eliminated. Then we compare $x$ with $a_{4i}$ and eliminate it, leaving a portion of an $(m-1)\times (n-2)$
monotone matrix, which can be searched using at most $(m-1)+(n-2)-1$ comparisons. In total, at most $m+n-2$
comparisons are needed.
\\[2.5mm]
\noindent \emph{Case 2.} The first row of $L$ has one more element than the second row. We can assume $|L| \ge 8$,
since for $|L|=4, 6,$ or 7 it is easy to verify that the first row of $L$ has two more elements than the second
row, which belong to Case 1. When $|L|\geq 8$, by Lemma \ref{lemma:L2} and \ref{lemma:L1} the second row of $L$
has two more elements than the third row, thus $L$ must have the following form
\begin{equation*}
\left[
\begin{array}{cccccccccc}
& \ldots & a_i   &a_{2i}&a_{4i} &a_{8i}& \\
& \ldots &a_{3i}&a_{6i}&a_{12i}&       & \\
& \ldots &a_{9i}&       &        &       & \\
& \ldots &       &       &        &       & \\
\end{array}%
\right]
\end{equation*}
First compare $x$ with $a_{4i}$.

If $x<a_{4i}$ then the two rightmost columns are known to be greater than $x$ and will be eliminated, leaving a
portion of an $m\times (n-2)$ monotone matrix, which can be searched using at most $m+(n-2)-1$ comparisons. In
total, at most $m+n-2$ comparisons are needed.

If $x>a_{4i}$ then all elements in the first row except $a_{8i}$ are known to be smaller than $x$ and will be
eliminated. Then compare $x$ with $a_{8i}$ and eliminate it, leaving a portion of an $(m-1)\times (n-1)$ monotone
matrix whose first row has two more elements than the second row. Thus it reduces to the situation of Case 1,
which needs at most $(m-1)+(n-1)-2$ comparisons. In total, at most $m+n-2$ comparisons are needed.
\\[2.5mm]
\noindent Therefore, in either case $m+n-2$ comparisons suffice.

\section{Proof of Lemma \ref{essential1}}

\label{appendix:essential1}

We prove the lemma by contradiction. Suppose that, under $RS_2$, $a_{j}\in E_n$ is not essential, i.e., there
exists an element $a_{i}\in A_n$ that cuts $a_{j}$, and $a_i$ is not in the unit containing $a_j$. We have the
following two cases.
\\
\\
%---------------------------------------------------Case 1.---------------------------------------------------------
\noindent \emph{Case 1.} $a_{i}$ and $a_{j}$ are in one layer. If $a_{i}$ does not belong to any unit, then $i\le
n/16$ and $a_{i}$ always cuts to the left up. If $a_i$ cuts an element $a_j$, then $2j\le i$, $32j\le 16i\le n$,
which implying that $a_{2j}$, $a_{4j}$, $a_{8j}$, $a_{16j}$ all exist in the row containing $a_{j}$, thus $a_j$
can not be in a unit. Therefore, $a_{i}$ can not cut any element in a unit. If $a_{i}$ belongs to a special unit,
4-$unit_s$, notice the form and the response strategy of special units, (\ref{R2:4-unit-s}), it is easy to see
that $a_{i}$ can not cut any element $a_{j}\in E_n$ in a different unit in the same layer. If $a_{i}$ belongs to a
general unit, we have the following eight subcases.
\begin{enumerate}
\item $a_{i}$ is the last element of a unit. Then $a_{i}$ always cuts to the
right bottom, $\{\ldots, \ulcorner a_i\}$, and $2i>n$. If $a_i$ cuts an element $a_j$, then $j\ge 2i>n$, which
contradicts with $j\le n$. Thus $a_{i}$ can not cut any element.

\item $a_{i}$ is the second last (first) element of a 2-unit, then it cuts to the left up, $\{a_{i} \lrcorner, a_{2i}\}$.
If $a_{i}$ cuts $a_{j}\in E_n$ in the same layer, then $a_{j}$ must be the first element of a row above $a_i$. In
addition, $a_{j}$ must be in the next row $R$ above $a_{i}$ (since by Lemma \ref{lemma:L2} the rows above $R$ have
at least 5 elements, if $a_j$ is in a unit in those rows then $a_j$ can not be the first element of that row). By
Lemma \ref{lemma:L1}, $R$ has 3 or 4 elements. If $R$ has 3 elements, then it is a 3-$unit_2$ and its first
element does not belong to $E_n$. If $R$ has 4 elements, then it is a 4-$unit_g$ and its first element does not
belong to $E_n$.

\item $a_{i}$ is the second last (second) element of a 3-$unit_1$, then it cuts to the right bottom,
$\{a_{i/2},\ulcorner a_{i}, a_{2i}\}$. Notice the form of 3-$unit_1$, (\ref{R2:3-unit-1}), $a_{i}$ can not cut any
element in a different unit in the same layer.

\item $a_{i}$ is the second last (second) element of a 3-$unit_2$, then it cuts to the left up,
$\{a_{i/2}, a_{i} \lrcorner, a_{2i}\}$. Since the row below $a_{i}$ has two elements, by Lemma \ref{lemma:L1} and
Lemma \ref{lemma:L2}, the next row above $a_{i}$, $R$, has 5 elements. Thus, if $a_{i}$ cuts $a_{j}\in E_n$ in the
same layer, $a_{j}$ must be in $R$, and is the first element of the 4-unit $U$ of $R$. Notice the form of special
units, (\ref{R2:4-unit-s}), in which the bottom row has one element. Therefore, $U$ is not a special unit and its
first element does not belong to $E_n$.

\item $a_{i}$ is the second last (third) element of a 4-$unit_g$, then it cuts to the right bottom,
$\{a_{i/4}, a_{i/2}, \ulcorner a_{i}, a_{2i}\}$. Denote by $U$ the 4-$unit_g$ which $a_{i}$ is in. Notice the form
of general units, (\ref{R2:4-unit-c}), $a_{j}$ must be the last element of $R$, where $R$ is the next row below
$a_{i}$ and has one less element than the row containing $a_i$. If $R$ has at least four elements, then $a_{j}$
must be the last element of a 4-$unit_g$, thus $a_{j}\notin E_n$. Otherwise $R$ must have exactly three elements,
then $a_{j}$ must be the third element of $R$. Since $R$ has one less element than the next row above it, by Lemma
\ref{lemma:L1} and Lemma \ref{lemma:L2}, the next row below $R$ has one element, it follows that $R$ is a
3-$unit_1$, thus its third element $a_{j}\notin E_n$.

\item $a_{i}$ is the third last (first) element of a 3-$unit$ (3-$unit_1$ or 3-$unit_2$), then it cuts to the left up,
$\{a_{i}\lrcorner, a_{2i}, a_{4i}\}$. If $a_{i}$ cuts $a_{j}\in E_n$ in the same layer, $a_j$ must be the first
element of $R$, where $R$ is the next row above $a_i$. In addition, $R$ must contain exactly four elements. Notice
the form of special units, (\ref{R2:4-unit-s}), $R$ is not a 4-$unit_s$, thus its first element $a_{j}\notin E_n$.

\item $a_{i}$ is the third last (second) element of a 4-$unit_g$, then it cuts to the left up,
$\{a_{i/2}, a_{i}\lrcorner, a_{2i}, a_{4i}\}$. In this case $a_j$ must be in the next row above $a_i$, $R$. In
addition, $R$ has one more element than the row containing $a_i$, and $a_j$ is the first element of the 4-unit $U$
in $R$. Notice the form of special units, (\ref{R2:4-unit-s}), $U$ is not a 4-$unit_s$, thus its first element
$a_{j}\notin E_n$.

\item $a_{i}$ is the fourth last (first) element of a 4-$unit_g$, then it cuts to the left up,
$\{a_{i}\lrcorner, a_{2i}, a_{4i}, a_{8i}\}$. If $a_i$ cuts an element $a_j$, then $2j\le i$, $16j\le 8i\le n$,
which implying that $a_{2j}$, $a_{4j}$, $a_{8j}$, $a_{16j}$ all exist in the row containing $a_{j}$, thus $a_j$
can not be in a unit. Therefore, $a_{i}$ can not cut any element in a unit.
\end{enumerate}

%---------------------------------------------------Case 2.---------------------------------------------------------
\noindent \emph{Case 2.}  $a_{i}$ and $a_{j}$ are in different layers. We first prove the following lemma that
will be useful later.

\begin{lemma} \label{at-least-five} For any $a_{j_1}, a_{j_2}\in A_n$ in different layers, if $j_1$ divides $j_2$,
then the quotient is at least 5.
\end{lemma}

\noindent \textbf{Proof of Lemma \ref{at-least-five}.} Suppose that $a_{j_1}\in L_{i_1}$ with $j_1={i_1\times
2^{k_1}\times 3^{s_1}}$, $a_{j_2}\in L_{i_2}$ with $j_2={i_2\times 2^{k_2}\times 3^{s_2}}$, where $L_{i_1}$ and
$L_{i_2}$ are different layers (i.e., $i_1\neq i_2$) and $j_1$ divides $j_2$. Since $i_1, i_2$ have no factor 2 or
3, we have ${k_1}\leq {k_2}$ and ${s_1}\leq {s_2}$, and $i_1$ divides $i_2$ with quotient at least 5. It follows
that $j_1$ divides $j_2$ with quotient at least 5.
\endpf

There are seven subcases in Case 2.

\begin{enumerate}
\item $a_{i}$ does not belong to any unit. Then $i\le \frac{n}{16}$ and $a_{i}$ always cuts to the
left up, by using the same argument at the beginning of Case 1, $a_{i}$ can not cut any element in a unit.

\item $a_{i}$ is the last element of a unit. Then $a_{i}$ always cuts to the right bottom,
$\{\ldots, \ulcorner a_i\}$, and $2i>n$. By using the same argument in subcase 1 of Case 1, $a_{i}$ can not cut
any element.

\item $a_i$ is the second last element of a
unit and cuts to the left up, $\{\ldots, a_i \lrcorner, a_{2i}\}$. By Lemma \ref{at-least-five}, if $a_i$ cuts
$a_{j}\in E_n$ in a different layer then $5j\leq i$, $10j\leq 2i\leq n$, thus $a_{2j}$, $a_{4j}$, $a_{8j}$ all
exist in the row containing $a_{j}$. If $a_j\in E_n$, $a_j$ can only be the first element of some special unit, it
follows that $j/2\in S_n=\{k\in I_n|\frac{n}{18}<k\le \frac{n}{16}\}$ (see Definition \ref{def:specialunit}), thus
$9j>n$, which contradicts with $10j\leq n$.

\item $a_{i}$ is the second last element of a unit and cuts to the right bottom, $\{\ldots, \ulcorner a_{i}, a_{2i}\}$.
Thus $4i>n$.  By Lemma \ref{at-least-five}, if $a_i$ cuts $a_{j}\in E_n$ in a different layer, then $j\ge 5i>n$,
which contradicts with $j\le n$.

\item $a_{i}$ is the third last element of a unit and
cuts to the left up, $\{ \ldots, a_{i} \lrcorner, a_{2i}, a_{4i}\}$. By Lemma \ref{at-least-five}, if $a_i$ cuts
$a_{j}\in E_n$ in a different layer then $5j\leq i$, thus $20j\leq 4i\leq n$. It follows that $a_{2j}$, $a_{4j}$,
$a_{8j}$, $a_{16j}$ all exist in the row containing $a_{j}$, thus $a_j$ can not be in a unit, which contradicts
with $a_{j}\in E_n$.

\item $a_i$ is the third last element of a unit and
cuts to the right bottom. In this case, $a_i$ must be the second element of a special unit, $\{a_{i/2},\ulcorner
a_{i}, a_{2i}, a_{4i}\}$. It follows that $i/4\in S_n=\{k\in I_n|\frac{n}{18}<k\le \frac{n}{16}\}$, thus
$i>\frac{2n}{9}$. However, by Lemma \ref{at-least-five}, if $a_i$ cuts $a_{j}\in E_n$ in a different layer then
$j\ge 5i>\frac{9i}{2}>n$, which contradicts with $j\le n$.

\item $a_{i}$ is the fourth last element of a unit, i.e., the first element of a 4-unit, $\{ a_{i}
\lrcorner, a_{2i}, a_{4i}, a_{8i}\}$. Then $a_{i}$ always cuts to the left up, and by using the same argument in
subcase 8 of Case 1, $a_{i}$ can not cut any element in a unit.
\end{enumerate}

\section{Proof of Lemma \ref{essential2}}

\label{appendix:essential2}

We prove the lemma by contradiction. Suppose that, under $RS^{*}_2$, $a_{j}\in E^{*}_n$ is not essential, i.e.,
there exists an element $a_{i}\in A_n$ that cuts $a_{j}$, and $a_i$ is not in the unit containing $a_j$. We have
the following two cases.
\\
\\
%---------------------------------------------------Case 1.---------------------------------------------------------
\noindent \emph{Case 1.} $a_{i}$ and $a_{j}$ are in one layer. If $a_{i}$ does not belong to any unit, then
$a_{i}$ always cuts to the left up, by using the same argument at the beginning of Case 1 in the proof of Lemma
\ref{essential1}, $a_{i}$ can not cut any element in a unit. If $a_{i}$ belongs to a special unit of the first or
the second class, i.e., a 4-$unit_{s,1}$ or 4-$unit_{s,2}$, notice the forms of these special units,
(\ref{R2:4-unit-s}) and (\ref{specialunit-2}), it is easy to see that $a_{i}$ can not cut any element $a_{j}\in
E^{*}_n$ in a different unit in the same layer (for the case where $a_{i}$ is in a 4-$unit_{s,2}$, notice that the
second row of any layer can not contain a special unit). For the case where $a_{i}$ belongs to a special unit of
the third class, i.e. a 4-$unit_{s,3}$, we first give two lemmas that will be useful. Similarly as Lemma
\ref{lemma:L2}, we have the following
\begin{lemma} \label{lemma:increment_two}
In any layer, there can not be four consecutive rows with lengths each increased by 2.
\end{lemma}

The correctness of Lemma \ref{lemma:increment_two} can be easily seen by noticing that in the third, fourth, fifth
and sixth row in (\ref{specialunit-3}), $9i$ exists in the layer since $9i<\frac{32i}{3}$.

\begin{lemma} \label{lemma:nonspecial}
If $R$ is the next row above or below a 4-$unit_{s,3}$ in the same layer, then $R$ does not contain a special
unit.
\end{lemma}

Lemma \ref{lemma:nonspecial} is true since by the definition of 4-$unit_{s,3}$, $R$ is not the first row of its
layer thus can not contain a 4-$unit_{s,1}$ or a 4-$unit_{s,2}$; and by Lemma \ref{lemma:increment_two}, $R$
either has one less element than the next row above $R$ or has one more element than the next row below $R$, thus
can not contain a 4-$unit_{s,3}$.

By Lemma \ref{lemma:nonspecial}, if $a_{i}$ belongs to a 4-$unit_{s,3}$, $a_{i}$ can not cut any element $a_{j}\in
E^{*}_n$ in a different unit in the same layer.
\\
\\
If $a_{i}$ belongs to a general unit, similarly as Case 1 in the proof of Lemma \ref{essential1}, we have the
eight subcases. For all these subcases we can show that, using the same arguments in Lemma \ref{essential1}
correspondingly (sometimes when comes to special units, we only need to replace them with the new defined special
units 4-$unit_{s,1}$, 4-$unit_{s,2}$ and 4-$unit_{s,3}$), $a_{i}$ can not cut $a_{j}\in E^{*}_n$.
\\
\\
%---------------------------------------------------Case 2.---------------------------------------------------------
\noindent \emph{Case 2.}  $a_{i}$ and $a_{j}$ are in different layers. We first give the following two lemmas that
will be useful later.

\begin{lemma} \label{at-least-seven}
For any $a_{j_1}, a_{j_2}\in A_n$ in different layers, if $j_1$ divides $j_2$, and $j_2$ is not divisible by 5,
then the quotient is at least 7.
\end{lemma}

\begin{lemma} \label{twelvegreater} If $i$ is the subscript of
a first element in a special unit 4-$unit_{s,1}$, 4-$unit_{s,2}$ or 4-$unit_{s,3}$, then $12i>n$.
\end{lemma}

Lemma \ref{at-least-seven} can be proved in a similar way as for Lemma \ref{at-least-five}, and Lemma
\ref{twelvegreater} is true since by the definitions of 4-$unit_{s,1}$'s, 4-$unit_{s,2}$'s and 4-$unit_{s,3}$'s,
the row containing a special unit always has two more elements than the next row below it.

Similarly as Case 2 in the proof of Lemma \ref{essential1}, we have the following seven subcases. Except subcase 3
and subcase 6, we can apply the same arguments in Lemma \ref{essential1} to all the following subcases
correspondingly.

\begin{enumerate}
\item $a_{i}$ does not belong to any unit. By using the same argument at the beginning of Case 1 in the proof of
Lemma \ref{essential1}, $a_{i}$ can not cut $a_{j}\in E^{*}_n$.

\item $a_{i}$ is the last element of a unit. Then $a_{i}$ always cuts to the
right bottom, $\{\ldots, \ulcorner a_i\}$. By using the same argument in subcase 1 of Case 1 in the proof of Lemma
\ref{essential1}, $a_{i}$ can not cut $a_{j}\in E^{*}_n$.

\item $a_i$ is the second last element of a
unit and cuts to the left up, $\{\ldots, a_i \lrcorner, a_{2i}\}$. By Lemma \ref{at-least-five}, if $a_i$ cuts
$a_{j}\in E^{*}_n$ in a different layer, we have $5j\leq i$, $10j\leq 2i\leq n$, thus $a_{2j}$, $a_{4j}$, $a_{8j}$
all exist in the row containing $a_{j}$. If $a_j\in E^{*}_n$, $a_j$ can only be the first element of a special
unit. We have two possibilities about $a_{i}$.
\begin{enumerate}
\item $a_{i}$ is the first element of a 2-unit or the second
element of a 3-$unit_2$. In this case $a_{j}$ must be the first element of a 4-$unit_{s,1}$, since otherwise in
the row containing $a_j$, there will be at least two elements before $a_j$, thus $a_i$ can not cut $a_j$. However,
by the form of 4-$unit_{s,1}$, (\ref{R2:4-unit-s}), if $a_{j}$ is the first element of a 4-$unit_{s,1}$, then
$9j>n$, this contradicts with $10j\leq n$. Thus this possibility is eliminated.

\item $a_{i}$ is the third element of a special unit. By the definitions of the new defined special units, $i$ is not
divisible by $5$. Thus by Lemma \ref{at-least-seven}, $7j\leq i$, $14j\leq 2i \leq n$. However, by Lemma
\ref{twelvegreater}, if $a_j$ is the first element of a special unit, then $12j>n$, which contradicts with $14j
\leq n$.
\end{enumerate}

\item $a_{i}$ is the second last element of a unit and cuts to the right bottom, $\{\ldots, \ulcorner a_{i}, a_{2i}\}$.
By using the same argument in subcase 4 of Case 2 in the proof of Lemma \ref{essential1}, $a_{i}$ can not cut
$a_{j}\in E^{*}_n$.

\item $a_{i}$ is the third last element of a unit and
cuts to the left up, $\{ \ldots, a_{i} \lrcorner, a_{2i}, a_{4i}\}$. By using the same argument in subcase 5 of
Case 2 in the proof of Lemma \ref{essential1}, $a_{i}$ can not cut $a_{j}\in E^{*}_n$.

\item $a_i$ is the third last element of a unit and cuts to the right bottom. In this case, $a_i$ must be the
second element of a special unit, $\{a_{i/2},\ulcorner a_{i}, a_{2i}, a_{4i}\}$. By the form of 4-$unit_{s,1}$,
(\ref{R2:4-unit-s}), $a_{i}$ can not be in a 4-$unit_{s,1}$, since otherwise we have $9\times i/2>n$, it follows
that $j\ge 5i>n$. Therefore, $a_{i}$ can only be the second element of a 4-$unit_{s,2}$ or 4-$unit_{s,3}$, thus
$i$ is divisible by 8. By Lemma \ref{twelvegreater}, $12\times i/2>n$, thus $6i>n$, since $j\ge 5i$ it must be the
case that $j=5i$. Thus $2j=10i>6i>n$, $a_{j}$ must be the last element of a unit. Since $i$ is divisible by 8,
$j=5i$ is also divisible by 8, thus $a_{j/2}$, $a_{j/4}$, $a_{j/8}$ all exist in the row containing $a_j$, it
follows that $a_j$ is the last element of a 4-unit. Therefore, if $a_{j}\in E^{*}_n$, $a_{j}$ must be the last
element of a special unit. By the definitions of the new defined special units, $j$ is not divisible by 5, which
contradicts with $j=5i$.

\item $a_{i}$ is the fourth last element of a unit, i.e., the first element of a 4-unit, $\{ a_{i}
\lrcorner, a_{2i}, a_{4i}, a_{8i}\}$. By using the same argument in subcase 8 of Case 1 in the proof of Lemma
\ref{essential1}, $a_{i}$ can not cut $a_{j}\in E^{*}_n$.
\end{enumerate}

\end{appendix}

\end{document}